\documentclass[twoside,english]{elsarticle}
\usepackage[LGR,T1]{fontenc}
\usepackage[latin9]{inputenc}
\usepackage{float}
\usepackage{units}
\usepackage{textcomp}
\usepackage{graphicx}

\makeatletter

\DeclareRobustCommand{\greektext}{%
  \fontencoding{LGR}\selectfont\def\encodingdefault{LGR}}
\DeclareRobustCommand{\textgreek}[1]{\leavevmode{\greektext #1}}
\DeclareFontEncoding{LGR}{}{}
\DeclareTextSymbol{\~}{LGR}{126}
\providecommand{\tabularnewline}{\\}
\floatstyle{ruled}
\newfloat{algorithm}{tbp}{loa}
\providecommand{\algorithmname}{Algorithm}
\floatname{algorithm}{\protect\algorithmname}

\journal{New Astronomy}

\usepackage{algorithm,algpseudocode}

\@ifundefined{showcaptionsetup}{}{%
 \PassOptionsToPackage{caption=false}{subfig}}
\usepackage{subfig}
\makeatother

\usepackage{babel}
\begin{document}

\begin{frontmatter}{}

\title{Variable Star Signature Classification using Slotted Symbolic Markov
Modeling}

\author[fit1]{K.B.~Johnston\corref{cor1}\corref{cor2}}

\ead{kyjohnst2000@my.fit.edu}

\author[fit2]{A.M.~Peter}

\ead{apeter@fit.edu}

\address[fit1]{Florida Institute of Technology, Physics and Space Sciences Department,
Melbourne, Florida, USA 32901}

\address[fit2]{Florida Institute of Technology, Systems Engineering Department ,
Melbourne, Florida, USA 32901}
\begin{abstract}
With the advent of digital astronomy, new benefits and new challenges
have been presented to the modern day astronomer. No longer can the
astronomer rely on manual processing, instead the profession as a
whole has begun to adopt more advanced computational means. This paper
focuses on the construction and application of a novel time-domain
signature extraction methodology and the development of a supporting
supervised pattern classification algorithm for the identification
of variable stars. A methodology for the reduction of stellar variable
observations (time-domain data) into a novel feature space representation
is introduced. The methodology presented will be referred to as Slotted
Symbolic Markov Modeling (SSMM) and has a number of advantages which
will be demonstrated to be beneficial; specifically to the supervised
classification of stellar variables. It will be shown that the methodology
outperformed a baseline standard methodology on a standardized set
of stellar light curve data. The performance on a set of data derived
from the LINEAR dataset will also be shown.\end{abstract}

\end{frontmatter}{}

\section{Introduction}

With the advent of digital astronomy, new benefits and new challenges
have been presented to the modern day astronomer. While data is captured
in a more efficient and accurate manor using digital means, the efficiency
of data retrieval has led to an overload of scientific data for processing
and storage. Where once the professional astronomer was faced with
ten to a hundred data points for a given night, the now more common
place \textquotedblleft full-sky survey\textquotedblright{} mission
results in millions of data points. This means that more stars, in
more detail are captured per night; but increasing data capture begets
exponentially increasing data processing. Database management, digital
signal processing, automated image reduction and statistical analysis
of data have all made their way to the forefront of tools for the
modern astronomer. Astro-statistics and astro-informatics are fields
which focus on the application and development of these tools to help
aid in the processing of large scale astronomical data resources. 

This paper focuses on one facet of this budding area, the construction
and application of a novel time-domain signature extraction methodology
and the development of a supporting supervised pattern classification
algorithm for the identification of variable stars. Given the reduction
of a survey of stars into a standard feature space, the problem of
using prior patterns to identify new observed patterns can be reduced
to time tested classification methodologies and algorithms. Such supervised
methods, so called because the user trains the algorithms prior to
application using patterns with known (hence the supervised) classes
or labels, provide a means to probabilistically determine the estimated
class type of new observations. These methods have two large advantages
over manual-classification procedures: the rate at which new data
is processed is dependent only on the computational processing power
available and the performance of a supervised classification algorithm
is quantifiable and consistent. Thus the supervised classification
algorithms produce rapid, efficient and consistent results.

A methodology for the reduction of stellar variable observations (time-domain
data) into a novel feature space representation is introduced. The
methodology presented will be referred to as Slotted Symbolic Markov
Modeling (SSMM) and has a number of advantages which will be demonstrated
over the course of this paper which are beneficial; specifically to
the supervised classification of stellar variables. The paper is structured
as follows. First, the data, prior efforts, and challenges uniquely
associated to classification of stars via stellar variability is reviewed.
Second, the novel methodology, SSMM, is outlined including the feature
space and signal conditioning methods used to extract the unique time-domain
signatures. Third, a set of classifiers (radial basis function neural
network, random forest/bagged decisions tree, k-nearest neighbor,
and Parzen window classifier) is trained and tested on the extracted
feature space using both a standardized stellar variability dataset
and the LINEAR dataset. Fourth, performance statistics is generated
for each classifier and a comparing and contrasting of the methods
are discussed. Lastly, an anomaly detection algorithm is generated
using the so called one-class Parzen Window Classifier and the LINEAR
dataset. The result will be the demonstration of the SSMM methodology
as being a highly competitive feature space reduction technique, for
usage in supervised classification algorithms.

\subsection{Related Work}

The idea of constructing a supervised classification algorithm for
stellar classification is not unique to this paper \citep{Dubath2011},
nor is the construction of a classifier for time variable stars. Methods
pursued include the construction of a detector to determine variability
\citep{Barclay2011}, the design of random forests for the detection
of photometric redshifts in spectra \citep{Carliles2010}, the detection
of transient events \citep{Djorgovski2012} and the development of
machine-assisted discovery of astronomical parameter relationships
\citep{Graham2013}. \citet{Debosscher2009} explored several classification
techniques for the supervised classification of variable stars, quantitatively
comparing the performance in terms of computational speed and performance.
Likewise, other efforts have focused on comparing speed and robustness
of various methods \citep{Blomme2011,Pichara2012,Pichara2013}. These
methods span both different classifiers and different spectral regimes,
including IR surveys \citep{Angeloni2014,Masci2014}, RF surveys \citep{Rebbapragada2011}
and optical \citep{Richards2012}. Methods for automated supervised
classification include procedures such as: direct parametric analysis
\citep{Udalski1999}, fully automated neural networking \citep{Pojmanski2000,Pojmanski2002}
and Bayesian classification \citep{Eyer2005}.

The majority of these references rely on periodicity domain feature
space reductions. \citet{Debosscher2009} and \citet{Templeton2004}
review a number of feature spaces and a number of efforts to reduce
the time domain data, most of which implement Fourier techniques,
primarily implementing the Lomb-Scargle (L-S) Method \citep{Lomb1976,Scargle1982},
to estimate the primary periodicity \citep{Eyer2005,Park2013,Richards2012,Ngeow2013,Deb2009}.
Lomb-Scargle is favored because of the flexibility it provides with
respect to observed datasets; it is frequently used when sample rates
are irregular and drop outs are common in the data being observed,
as is often the case with astronomical observations. \citet{Long2014}
advance L-S even further, introducing multi-band (multidimensional)
generalized L-S, allowing the algorithm to take advantage of information
across filters, in cases where multi-channel time-domain data is available.
There have also been efforts to estimate frequency using techniques
other than L-S such as the Correntropy Kernelized Periodogram, \citep{Huijse2011}
or MUlti SIgnal Classificator \citep{Tagliaferri2003}.

The assumption of the light curve being periodic, or even that the
functionality of the signal being represented in the limited Fourier
space that Lomb-Scargle uses, has been shown \citep{Palaversa2013,Barclay2011}
to result in biases and other challenges when used for signature identification
purposes. Supervised classification algorithms implementing these
frequency estimation algorithms do so to generate an estimate of primary
frequency; the primary frequency is then used to fold all observations
resulting in a plot of magnitude vs. phase, something \citet{Deb2009}
refer to as \textquotedblleft reconstruction\textquotedblright . After
some interpolation to place the magnitude vs. phase plots on similar
regularly sampled scales, the new folded time series can be directly
compared (1-to-1) with known folded time series. Comparisons can be
performed via distance metric \citep{Tagliaferri2003}, correlation
\citep{Protopapas2006}, further feature space reduction \citep{Debosscher2009}
or more novel methods \citep{Huijse2012}. It should be noted that
the family of stars with the label \textquotedblleft stellar variable\textquotedblright{}
is a large and diverse population: eclipsing binaries, irregularly
pulsating variables, nova (stars in outburst), multi-model variables,
and many others are frequently processed using the described methods
despite the underlying stellar variability functionality not naturally
lending itself to Fourier decomposition and the associated assumptions
that accompany the said decomposition. Indeed this is why \citet{Szatmary1994,Barclay2011,Palaversa2013}
and others suggest using other decomposition methods such as discrete
wavelet transformations, which have been shown to be powerful in the
effort to decompose a time series into the time-frequency (phase)
space for analysis \citep{Torrence1998}. It is noted that the possibilities
beyond Fourier domain analysis time series comparison are too numerous
to outline here; for those who are interested, the near complete review
by \citet{Fulcher2013} is highly recommended.

\subsection{Data Specific Challenges}

The classification of time series data has a number of considerations
that need to be made. In this section, we detail computational issues
associated with processing astronomical time series and propose appropriate
techniques to mitigate the challenges.

\subsubsection{Continuous Time Series Data}

Stellar variable time series data can roughly be described as passively
observed time series snippets, extracted from what is a contiguous
signal (star shine) over multiple nights or sets of observations.
The continuous nature of the time series provides both complications
and opportunities for time series analysis. The time series signature
have the potential to change over time, and new observations mean
increased opportunity for an unstable signature over the long term.
If the time signature does not change, then new observations will
result in additive information that will be used to further define
the signature function associated with the class. Implementing a methodology
that will address both issues (potential for change and potential
for additional information) would be beneficial. If the sampling was
regular (and continuous) Short-Time Fourier Transforms (Spectrograms)
or Periodiograms would be ideal, although these methods would be complicated
to turn into, or extract from, the signature pattern of the variable
star as the dimensions of the spectrogram would grow with increasing
time observations. Likewise, the data analyzed cannot be necessarily
represented in Fourier space (perfectly) and while the wavelet version
of the spectrogram or scalogram \citep{Rioul1991,Szatmary1994,Bolos2014}
could be used, the data is also irregularly sampled further complicating
the analysis. Methods for obtaining regularly spaced samples from
irregular samples are known \citep{Rehfeld2014,Broersen2009}, however,
these methods have unforeseen effects on the frequency domain signature
which is being extracted, thereby corrupting the signature pattern.

\subsubsection{Irregular Sampling}

Astronomical time series data is also frequently irregular, i.e.,
there is no associated fixed $\Delta t$ over the whole of the data
that is consistent with the observation. Even when there is a consistent
observation rate, this rate is often broken up because of a given
observational plan, day-light interference or weather related constraints.
Whatever method is used must be able to handle various irregular sampling
rates and observational dropouts, without introducing biases and artifacts
into the derived feature space that will be used for classification.
Most analysis methods require or at least depend on regularized samples.
Those that do not, either require some form of transformation from
irregular to regular sample rate by a defined methodology, or apply
some assumption about the time-domain function that generated the
variation to begin with (such as L-S). Irregular Sampling solutions
\citep{Bos2002,Broersen2009} to address this problem, can be defined
one of three ways: Slotting Methods which model points along the time
line using fuzzy or hard models \citep{Rehfeld2011,Rehfeld2014},
re-sampling estimators which use interpolation to generate the \textquotedblleft missing
points\textquotedblright{} and obtain a consistent sample rate, and
L-S like estimators which apply a model or basis function across the
time series and maximizes the coefficients of the basis function to
find an accurate representation of the time series.

\subsubsection{Signature Representations}

The stellar variable moniker covers a wide variety of variable types:
stationary (consistently repeating identical patterns), non-stationary
(patterns that increase/decrease in frequency over time), non-regular
variances (variances that change over the course of time, shape changes),
as well as both Fourier and non-Fourier sequences/patterns. Pure time-domain
signals do not lend themselves to signature identification and pattern
matching, as their domain is infinite in terms of potential discrete
data (dimensionality). So not only must a feature space representation
be found, but the dimensionality should not increase with increasing
data. There are a number of time-domain dimensionality reduction methodologies
available, DFT and DWT are two of the big contenders in today\textquoteright s
research. Piecewise Aggregate Approximation \citep{Keogh2001} and
Symbolic Aggregate Approximation \citep{lin2007experiencing} methodologies
however, has been shown to compete with both methods\citep{Lin2012},
and in some cases has been shown to perform better when pattern matching
is of interest (and not necessarily determination of frequency or
underlying features of the generating time domain signal).

\section{Proposed Feature Extraction Methodology}

The algorithm designed encompasses the analysis, reduction and classification
of data. The a priori distribution of class labels are roughly evenly
distributed for both studies, therefore the approach uses a multi-class
classifier. Should the class labels with additional data become unbalanced,
other approaches are possible \citep{Rifkin2004}. Based on the outlined
data/domain specific challenges, this paper will attempt to develop
a feature space extraction methodology that will construct an analysis
of stellar variables and characterize the shape of the periodic stellar
variable signature. A number of methods have been demonstrated that
fit this profile \citep{Grabocka2012,Fu2011,Fulcher2013}, however
many of these methods focus on identifying a specific time series
shape sequence in a long(er) continuous time series, and not necessarily
on the differentiation between time series sequences. To address these
domain specific challenges, the following methodology outline is implemented: 
\begin{enumerate}
\item To address the irregular sampling rate, a slotting methodology is
used \citep{Rehfeld2011}: Gaussian kernel window slotting with overlap.
The slotting methodology is used to generate estimates of amplitudes
at regularized points, with the result being a up-sampled conditioned
waveform. This has been shown to be useful in the modeling and reconstruction
of variability dynamics\citep{Rehfeld2014}, and is similar to the
methodologies used to perform Piecewise Aggregate Approximation \citep{Keogh2001}.
\item To reduce the conditioned time series into a usable feature space,
the amplitudes of the conditioned time series will be mapped to a
discrete state space based on a standardized alphabet. The result
is the state space representation of the time domain signal, and is
similar to the methodologies used to perform Symbolic Aggregate Approximation
\citep{lin2007experiencing}.
\item The state space transitions are then modeled as a first order Markov
Chain, and the state transition probability matrix (Markov Matrix)
is generated, a procedure unique to this study. It will be shown that
a mapping of the transitions from observation to observation will
provide an accurate and flexible characterization of the stellar variability
signature. 
\end{enumerate}
The Markov Matrix is unfolded into a vector, and is the signature
pattern (feature vector) used in the classification of time-domain
signals for this study.

\subsection{Slotting (Irregular Sampling)}

Each waveform is modeled using the slotting re-sampling methodology
for irregularly sampled waveforms outlined in \citet{Rehfeld2011}.
The slotting method results in a set of regularly sampled amplitude
estimates; these are the conditioned waveforms for this analysis.
Let the set of $\{y(t_{n})\}_{n=1}^{N}$ samples, where $t_{1}<t_{2}<t_{3}<...<t_{N}$
and there are N samples, be the initial time series dataset. The observed
time series data is standardized (subtract the mean, divide by the
standard deviation), and then the slotting procedure is applied. If
$x[i]\gets{y(t_{i})}_{i=1}^{N}$, then the algorithm to generate the
slotted time domain data is given in Algorithm \ref{Algorithm1}.

\begin{algorithm}[H]
\protect\caption{Gaussian Kernel Slotting\label{Algorithm1}}

\begin{algorithmic}[1] 
\Procedure{GaussianKernelSlotting}{$x[i], t[i], w, \lambda$}
\State 
\State $x_{prime}[i] \gets (x[i] - mean(x[i]))/std(x[i])$ \Comment{Standardize Amplitudes}
\State $t[i] \gets t[i] - min(t[i])$  \Comment{Start at Time Origin}
\State $slotCenters \gets 0:\frac{w}{4}:max(t[i]) + w$  \Comment{Make Slot Locations} 
\State $timeSeriesSets = []$  \Comment{Initialize Time Series Sets} 
\State $slotSet = []$  \Comment{Make an Empty Slot Set} 
\State 
\While{$i < length(slotCenters)$} \Comment{Compute Slots}
	\State $idx \gets $ all $t$ in interval $[slotCenters - w, slotCenters + w]$
	\State $inSlotX \gets x[idx]$ 
	\State $inSlotT \gets t[idx]$ 
	\State 
	\If{$inSlot$ is empty} \Comment{There is a Gap}
		\If{$slotSet$ is empty} \Comment{Move to Where Data is}
			\State $currentPt \gets$ find next $t > slotCenters + w$ 
			\State $i \gets$ find last $slotCenters < t[currentPt]$
		\Else \Comment{Store the Slotted Estimates}
			\State add $slotSet$ to structure $timeSeriesSets$ 
			\State $slotSet \gets [ ]$ 
		\EndIf 
	\Else 
		\State $weights \gets exp(-((inSlot-slotCenters)^2*\lambda))$
		\State $meanAmp \gets sum(weights*inSlotX)/sum(weights)$ 
		\State add $meanAmp$ to the current slotSet
	\EndIf
	\State $i++$ 
\EndWhile\label{slottingendwhile} 
\EndProcedure 
\end{algorithmic}
\end{algorithm}

The slotting procedure selects a set of points about a point on the
time grid and within the slot to be considered; for this implementation
an overlapping slot (75\% overlap) was used. Where overlapping here
means that the window width is larger then the distance between slot
centers. These points are then weighted using a Gaussian model to
generate a weighted mean amplitude for the slot. This methodology
is effectively Kernel Smoothing with Slotting \citep{Li2007}. The
time series, with irregular sampling and large gaps is conditioned
by the Gaussian slotting method. Gaps in the waveform are defined
as regions where a slot contains no observations. Continuous observations
(segments) are the set of observations between the gaps. This results
in a set of waveforms that have equally spaced sampling. This conditioning
is also similar to the Piecewise Aggregation Approximation\citep{Lin2003,Keogh2001}.
Instead of down-sampling the time domain datasets as PAA does however,
the data is up-sampled using the slotting methodology. This is necessary
because of the sparsity of the time domain sampling of astronomical
data.

\subsection{State Space Representation}

\begin{figure}[H]
\protect\caption{Example State Space Representation \label{fig:StateSpaceRep}}

\centering{}\includegraphics[scale=0.5]{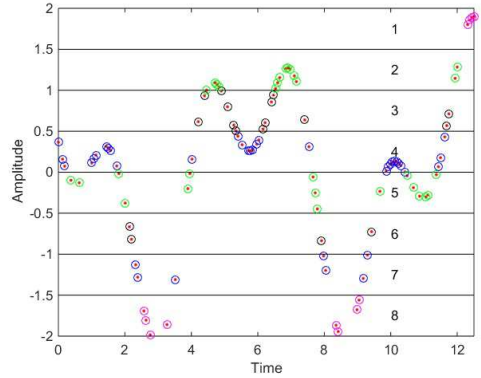}
\end{figure}

If it is assumed that the conditioned standardized waveform segments
have an amplitude distribution that approximates a Gaussian distribution
(which they won\textquoteright t, but that is irrelevant to the effort),
then using a methodology similar to Symbolic Aggregate Approximation
\citep{lin2007experiencing,Lin2012} methodologies, an alphabet (state
space) is defined based on our assumptions as an alphabet extending
between \textpm 2\textgreek{sv} and will encompass 95\% of the amplitudes
observed. This need not always be the case, but the advantage of the
standardization of the waveform is that, with some degree of confidence
the information from the waveform is contained roughly between \textpm 2\textgreek{sv}.
The resolution of the alphabet granularity is to be determined via
cross-validation to determine an optimal resolution. Figure \ref{fig:StateSpaceRep}
demonstrates a eight state translation; the alphabet will be significantly
more resolved then this for astronomical waveforms.

The set of state transitions, the transformation of the conditioned
signal, is used to populate a transition probability matrix or first
order Markov Matrix.

\subsection{Transition Probability Matrix (Markov Matrix)}

The transition state frequencies are estimated for signal measured
between empty slots, transitions are not evaluated between day-night
periods, or between slews (changes in observation directions during
a night) and only evaluated for continuous observations. Each continuous
set of conditioned waveforms (with Slotting and State Approximation
applied) is used to populate the empty matrix $P$, with dimensions
equal to $r\times r$, where $r$ is the number of states, is built.
The matrix is populated using the following rules:
\begin{itemize}
\item $N_{ij}$is the number of observation pairs $x[t]$ and $x[t+1]$
with $x[t]$ is state $s_{i}$ and $x[t+1]$ in state $r_{j}$
\item $N_{i}$is the number of observation pairs $x[t]$ and $x[t+1]$ with
$x[t]$ in state $s_{i}$ and $x[t+1]$ in any one of the states $j=1,...,r$
\end{itemize}
The now populated matrix $P$ is a transition frequency matrix, with
each row $i$ representing a frequency distribution (histogram) of
transitions out of the state $s_{i}$. The transition probability
matrix is approximated by converting the elements of P by approximating
the transition probabilities using $P_{ij}=\nicefrac{N_{ij}}{N_{i}}$.
The resulting matrix is often described as a first order Markov Matrix
\citep{Ross2013}. State changes are based on only the observation-to-observation
amplitude changes; the matrix is a representation of the linearly
interpolated sequence \citep{Ge2000}. Furthermore, the matrix is
unpacked similar to image analysis methods into a feature space vector,
with dimensions depend on the resolution and bounds of the states.
The algorithm to process the time-domain conditioned data is given
in Algorithm \ref{Algorithm2}.

\begin{algorithm}[H]
\protect\caption{Markov Matrix Generation \label{Algorithm2}}

\begin{algorithmic}[2] 
\Procedure{MarkovMatrixGeneration}{$timeSeriesSets,s$} 
\State $markovMatrix = []$ 
\For{$i:= 1$ to length of $timeSeriesSets$}
	\State $markovMatrixPrime \gets []$
	\State $currentSlotSet \gets markovMatrixPrime[j]$
	\For{$k:= 1$ to length of $currentSlotSet$}
		\State $idxIn  \gets $ find state containing $currentSlotSet[k-1]$ 
		\State $idxOut \gets $ find state containing $currentSlotSet[k]$ 
		\State $markovMatrixPrime[idxIn, idxOut]++$
	\EndFor
	\State $markovMatrix \gets markovMatrix + markovMatrixPrime$
\EndFor
\State $N_i =$ sum along row of $markovMatrix$
\For{$j := 1$ to length of $s$}
	\If{$N_i \neq 0$}
		\State $markovMatrix[:,j] \gets \frac{N_{ij}}{N_i}$			\Comment{Estimate Markov Matrix}
	\EndIf
\EndFor
\EndProcedure 
\end{algorithmic}
\end{algorithm}

The resulting Markov Matrix is unpacked into a feature vector given
by:

\begin{equation}
\mathbf{P_{\mathit{i}}}=\left[\begin{array}{cccc}
p_{11} & p_{12} & ... & p_{1r}\\
p_{21} & p_{22} & \cdots & \cdots\\
\vdots & \vdots & \ddots & \vdots\\
p_{r1} & p_{21} & ... & p_{rr}
\end{array}\right]\Rightarrow x_{i}=\left[\begin{array}{cccccc}
p_{11} & p_{12} & ... & p_{21} & ... & p_{rr}\end{array}\right]
\end{equation}

Where $\mathbf{P}_{i}$ is the Markov Chain of the $i^{th}$ input
training set, and $x_{i}$ is the $i^{th}$ input unfolded training
pattern.

\subsection{Feature Space Reduction (ECVA)}

The resolution of the state set needs to be small to avoid loss of
information resulting from over generalization. However, if the state
resolution is too small the sparsity of the transition matrix will
result in a shape signature that is too dependent on noise and the
\textquotedblleft individualness\textquotedblright{} of specific waveform
to be of any use. Thus additional processing is necessary for further
analysis; even a small set of states (12 x 12) will result in a feature
vector with high dimensionality (144 dimensions). While a window and
overlap size is assumed for the slotting to address the irregular
sampling of the time series data, there are two adjustable features
associated with this analysis: the kernel width associated with the
slotting and the state space (alphabet) resolution. It is apparent
that a range of resolutions and kernel width need to be tested to
determine best performance given a generic supervised classifier.
For these purposes a rapid initial classification algorithm, General
Quadratic Discriminate Analysis \citep{Duda2012}, was implemented
to estimate the mis-classification rate (wrong decisions/total decisions).
Not all states will be observed, i.e. the high dimensional feature
vector will have information contained in a small subset of elements.
Dimensionality reduction methods are often necessary for implementation
of classification algorithms, in particular QDA where the construction
of a covariance matrix of a sparse feature space can be problematic.

The reduction of the large, sparse, feature vector resulting from
the unpacking of the Markov Matrix is performed via extended canonical
variate analysis or ECVA \citep{Norgaard2006}. The methodology for
ECVA has roots in principle component analysis (PCA). PCA is a procedure
performed on large multidimensional datasets with the intent of rotating
what is a set of possibly correlated dimensions into a set of linearly
uncorrelated variables \citep{Sch06}. The transformation results
in a dataset, where the first principle component (dimension) has
the largest possible variance. PCA is an unsupervised methodology,
i.e. a priori known labels for the data being processed is not taken
into consideration, thus a reduction in feature dimensionality and
while it maximizes the variance it might not maximize the linear separability
of the class space. In contrast to PCA, Canonical Variate Analysis
does take class labels into considerations. The variation between
groups is maximized resulting in a transformation that benefits the
goal of separating classes. Given a set of data $\mathbf{x}$ with:
$g$ different classes, $n_{i}$ observations of each class, and $r\times r$
dimensions in each observation; following \citet{Johnson1992}, the
within-group and between-group covariance matrix is defined as:

\begin{equation}
\mathbf{S}_{within}=\frac{1}{n-g}\sum_{i=1}^{g}\sum_{j=1}^{n_{i}}(\mathbf{x}_{ij}-\bar{\mathbf{x}}_{ij})(\mathbf{x}_{ij}-\bar{\mathbf{x}}_{i})'
\end{equation}

\begin{equation}
\mathbf{S}_{between}=\frac{1}{g-1}\sum_{i=1}^{g}n_{i}(\mathbf{x}_{i}-\bar{\mathbf{x}})(\mathbf{x}_{i}-\bar{\mathbf{x}})'
\end{equation}

where $n=\sum_{i=1}^{g}n_{i}$, $\bar{\mathbf{x}}_{i}=\frac{1}{n_{i}}\sum_{j=1}^{n_{i}}\mathbf{x}_{ij}$,
and $\bar{\mathbf{x}}=\frac{1}{n}\sum_{j=1}^{n_{i}}n_{i}\mathbf{x}_{i}$.
CVA attempts to maximize the function:

\begin{equation}
J(\mathbf{w})=\frac{\mathbf{w}'\mathbf{S}_{between}\mathbf{w}}{\mathbf{w}'\mathbf{S}_{within}\mathbf{w}}
\end{equation}

Which is solvable so long as $\mathbf{\mathbf{S}_{\mathit{within}}}$is
non-singular, which need not be the case, especially when analyzing
multi-collinear data. When the case arises that the dimensions of
the observed patterns are multi-collinear additional considerations
need to be made. \citet{Norgaard2006} outlines a methodology for
handling these cases in CVA; the equation $\mathbf{S}_{between}\mathbf{w}=\lambda\mathbf{w}\mathbf{S}_{within}$
is reformulated (in the two class case) as: $(\bar{\mathbf{x}}_{1}-\bar{\mathbf{x}}_{2})(\bar{\mathbf{x}}_{1}-\bar{\mathbf{x}}_{2})'\mathbf{w}=\lambda\mathbf{w}\mathbf{S}_{within}$,
it is then shown that $(\bar{\mathbf{x}}_{1}-\bar{\mathbf{x}}_{2})'\mathbf{w}$
is a scalar value, and so the equation is rewritten in linear form
as $\mathbf{y}=\mathbf{Rb}+\mathbf{f}$ where $\mathbf{R}=\mathbf{S}_{within}$and
$\mathbf{b}=\mathbf{w}$. Likewise for the multi-group case ($g>2$)
this methodology can be expanded, by having $\mathbf{y}$ contain
as columns the differences between each group mean and the overall
mean. Partial least squares analysis, PLS2 \citep{Wold1939}, is used
to solve the above linear equation, resulting in an estimate of $\mathbf{w}$,
and given that, an estimate of the canonical variates (the reduced
dimension set). ECVA is applied to the set of patterns and labels,
a corresponding feature space that is of dimension $n$ by $g-1$
is constructed.

\section{Implementation of Methodology}

\subsection{Datasets}

Two datasets are addressed here, the first is the STARLIGHT dataset
from the UCR time series database, the second is published data from
the LINEAR survey. The UCR time series dataset is used to base line
the time-domain dataset feature extraction methodology proposed, it
is compared to the results published on the UCR website. The UCR time
series data contains only time domain data that has already been folded
and put into magnitude phase space, no differential photometric data
from either SDSS or 2MASS, nor star identifications for these data,
could be recovered, and only three class types are provided which
are not defined besides by number. The second dataset, the LINEAR
survey, provides an example of a modern large scale astronomical survey,
contains time-domain data that has not been folded or otherwise manipulated,
is already associated with SDSS and 2MASS photometric values, and
has five identified stellar variable types. For each dataset, the
state space resolution and the kernel widths for the slotting methods
will be optimized using 5-fold cross-validation. The performances
of four classifiers on only the time-domain dataset for the UCR data,
and on the mixture of time-domain data and differential photometric
data for the LINEAR survey, are estimated using 5-fold cross-validation
and testing. The performances of the classifiers will be compared.
Finally an anomaly detection algorithm will be trained and tested,
for the LINEAR dataset.

\subsection{Pattern Classification Algorithm}

The training set is used for 5-Fold cross-validation, and a set of
four classification algorithms are tested \citep{Hastie2009,Duda2012}:
k Nearest Neighbor (k-NN), Parzen Window Classifier (PWC), Radial
Basis Function Neural Network (RBF-NN), and Random Forest (RF). Cross-validation
is used to determine optimal classification parameters (e.g., kernel
width) for each of the classification algorithms. The first three
algorithms implemented were designed by the authors in MATLAB, based
on \citet{Duda2012} (k-NN and PWC) and \citet{Hastie2009} (RBF-NN)
algorithm outlines. Request for the implemented code should be made
to the authors directly.

\subsubsection{k-NN}

The k nearest neighbor algorithm is a non-parametric classification
method; it uses a voting scheme based on an initial training set to
determine the estimated label. For a given new observation, the $L_{2}$
Euclidean distance is found between the new observation and all points
in the training set. The distances are sorted, and the k closest training
sample labels are used to determine the new observed sample estimated
label (majority rule). Cross-validation is used to find an optimal
k value, where k is any integer greater than zero.

\subsubsection{PWC}

Parzen windows classification is a technique for non-parametric density
estimation, which is also used for classification \citep{Parzen1962,Duda2012}.
Using a given kernel function, the technique approximates a given
training set distribution via a linear combination of kernels centered
on the observed points. As the PWC algorithm (much like a k-NN) does
not require a training phase, as the data points are used explicitly
to infer a decision space. Rather than choosing the k nearest neighbors
of a test point and labeling the test point with the weighted majority
of its neighbor's votes, one can consider all points in the voting
scheme and assign their weight by means of the kernel function. With
Gaussian kernels, the weight decreases exponentially with the square
of the distance, so far away points are practically irrelevant. Cross-validation
is necessary however, to determine an optimal value of h, the \textquotedblleft width\textquotedblright{}
of the radial basis function (or whatever kernel is being used).

\subsubsection{RBF-NN}

A radial basis function neural network (RBF-NN) classification scheme
is used to generate a classifier. Using RBF-NN, the observed patterns
are first transformed into a new high-dimensional space. The RBF-NN
relies on the transformation of the data provided (measured) using
the kernel (radial basis) function. These kernels are representative
of the measured data and are often generated using prior knowledge.
The kernel function used is dependent on the prior knowledge available,
which for our classifier is means generated based on the input data
points. Each observation with dimension D is translated using the
individual Kernels. Thus if there are 100 individual observations,
the transformation for a given measurement vector will be a resulting
vector of 100. Alternatively, k-mean clustering could be used to reduce
the individual datasets to a representative kernel set allowing for
the resolution of the kernel transformation, but reducing the number
of computations necessary. Each dimension then is no longer a measurement,
but a distance between the measurements to the training data. After
the transformation of the data from the observed set to the RBF the
data is passed to the LRC algorithm. The logistic regression model
arises from the desire to model the posterior probability of the K
classes via linear functions in x, while at the same time ensuring
that they sum to one and remain in the range $[0,1]$.

\subsubsection{Random Forest Classifier}

To generate the random forest classifier, the TreeBagger algorithm
in MATLAB is implemented. The algorithm generates $n$ decision trees
on the provided training sample. The $n$ decision trees operate on
any new observed pattern, and the decision made by each tree are conglomerated
together (majority rule) to generate a combined estimated label. To
generate Breiman's 'random forest' algorithm \citep{Breiman1984},
the value NVarToSample is provided a value (other than \textquoteleft all\textquoteright )
and a random set of variables is used to generate the decision trees;
see the MATLAB TreeBagger documentation for more information.

\subsection{Comparison to Standard Set (UCR)}

\begin{figure}[H]
\protect\caption{UCR Phased Light Curves. Classes are given by number only: 1 = Blue
Line, 2 = Green Small Dashed Line, 3 = Red Big Dashed Line\label{fig:UCR-Phased-Light} }

\includegraphics[scale=0.5]{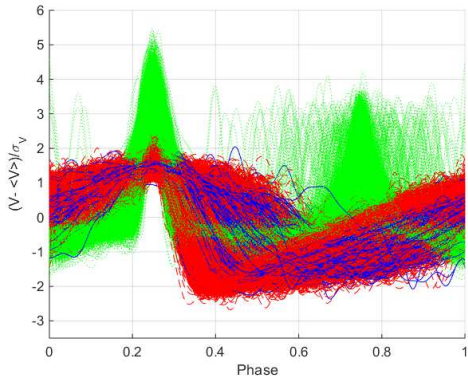}

\end{figure}

The UCR time domain datasets are used to basis classification methodologies
\citep{Keogh2011}. The UCR time domain datasets \citep{Protopapas2006},
are derived from a set of Cepheid, RRLyrae, and Eclipsing Binary Stars.
The time-domain datasets have been phased (folded) via the primary
period and smoothed using the SUPERSMOOTHER algorithm \citep{Reimann1994}
by the Protopapas study prior to being provided to the UCR database.
The waveforms received from UCR are amplitude as a function of phase;
the SUPERSMOOTHER algorithm was also used \citep{Protopapas2006}
to produce regular samples (in the amplitude vs. phase space). The
sub-groups of each of the three classes are combined together in the
UCR data (i.e., RRab + RRc = RR), similarly the data is taken from
two different studies (OGLE and MACHO). A plot of the phased light
curves is given in Figure \ref{fig:UCR-Phased-Light}.

Class analysis is a secondary effort when applying the methodology
outlined to the UCR dataset, the primary concern is a demonstration
of performance of the supervised classification methodology with respect
to the baseline performance reported by UCR implementing a simple
waveform nearest neighbor algorithm.

\subsubsection{Analysis}

\begin{figure}[H]
\protect\caption{ECVA reduced feature space using the UCR Star Light Curve Data\label{fig:ECVA_UCR}}

\centering{}\includegraphics[scale=0.5]{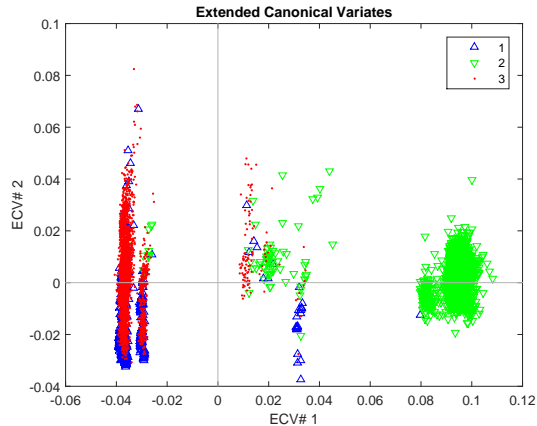}
\end{figure}

The folded waveforms are treated identical to the unfolded waveforms
in terms of the processing presented. Values of phase were generated
to accommodate the slotting technique, thereby allowing the functionally
developed to be used for both amplitude vs. time (LINEAR) as well
as amplitude vs. phase (UCR). The slotting, State Space Representation,
Markov Matrix and ECVA flow is implemented exactly the same. As there
are only three classes in the dataset, the ECVA algorithm results
in a dimensionality of only two ($g-1$). There is no accompanying
differential photometric data with the time-domain data, so only the
time-domain data will be focused on for this analysis. The resulting
ECVA plot is presented in Figure \ref{fig:ECVA_UCR}.

\begin{figure}[H]
\protect\caption{Classifier Optimization for UCR Data}

\begin{centering}
\subfloat[Nearest Neighbor Classifiers\label{fig:kNN_UCR}]{

\protect\includegraphics[scale=0.5]{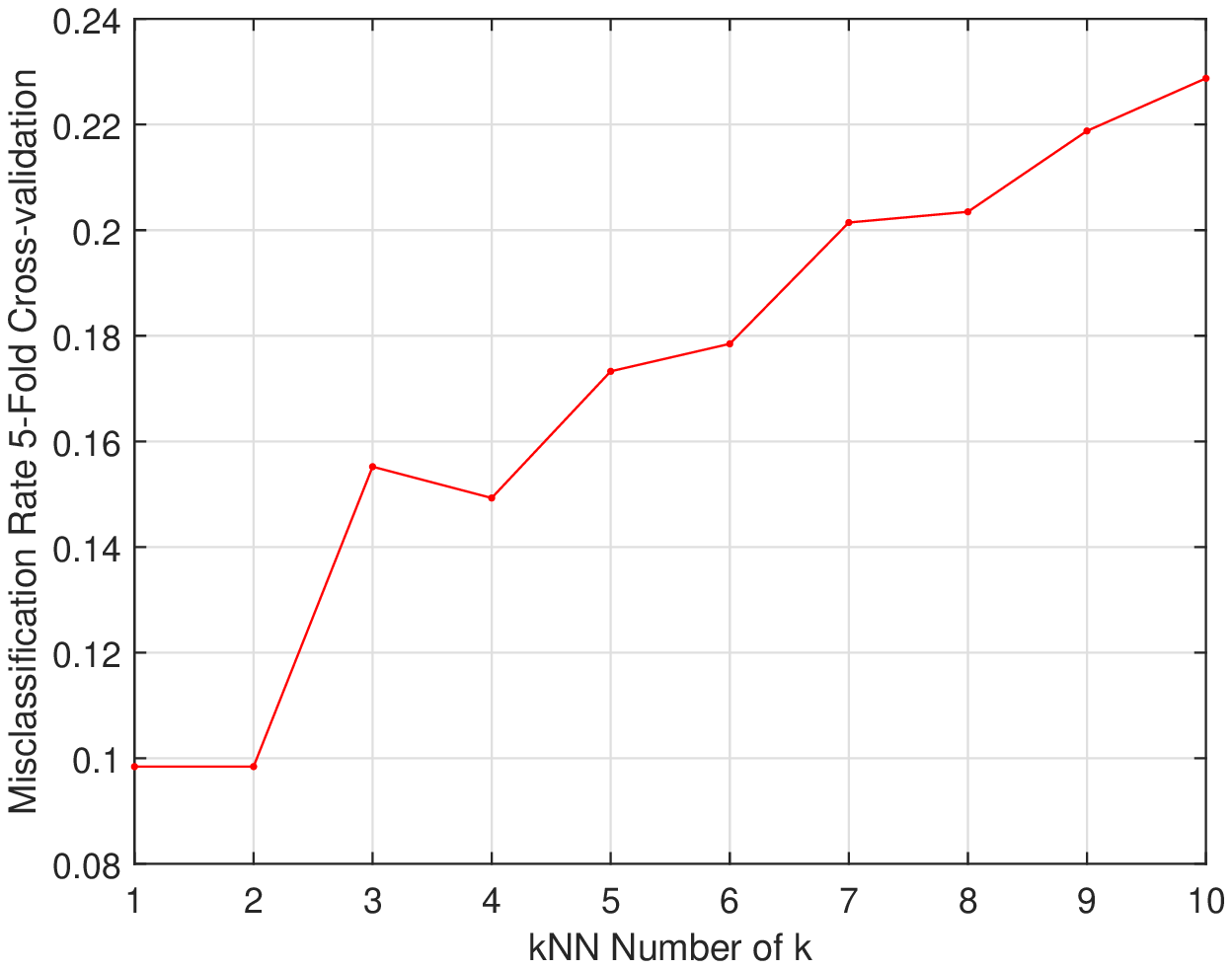}

}\subfloat[Parzen Window Classifier\label{fig:PWC_UCR}]{

\protect\includegraphics[scale=0.5]{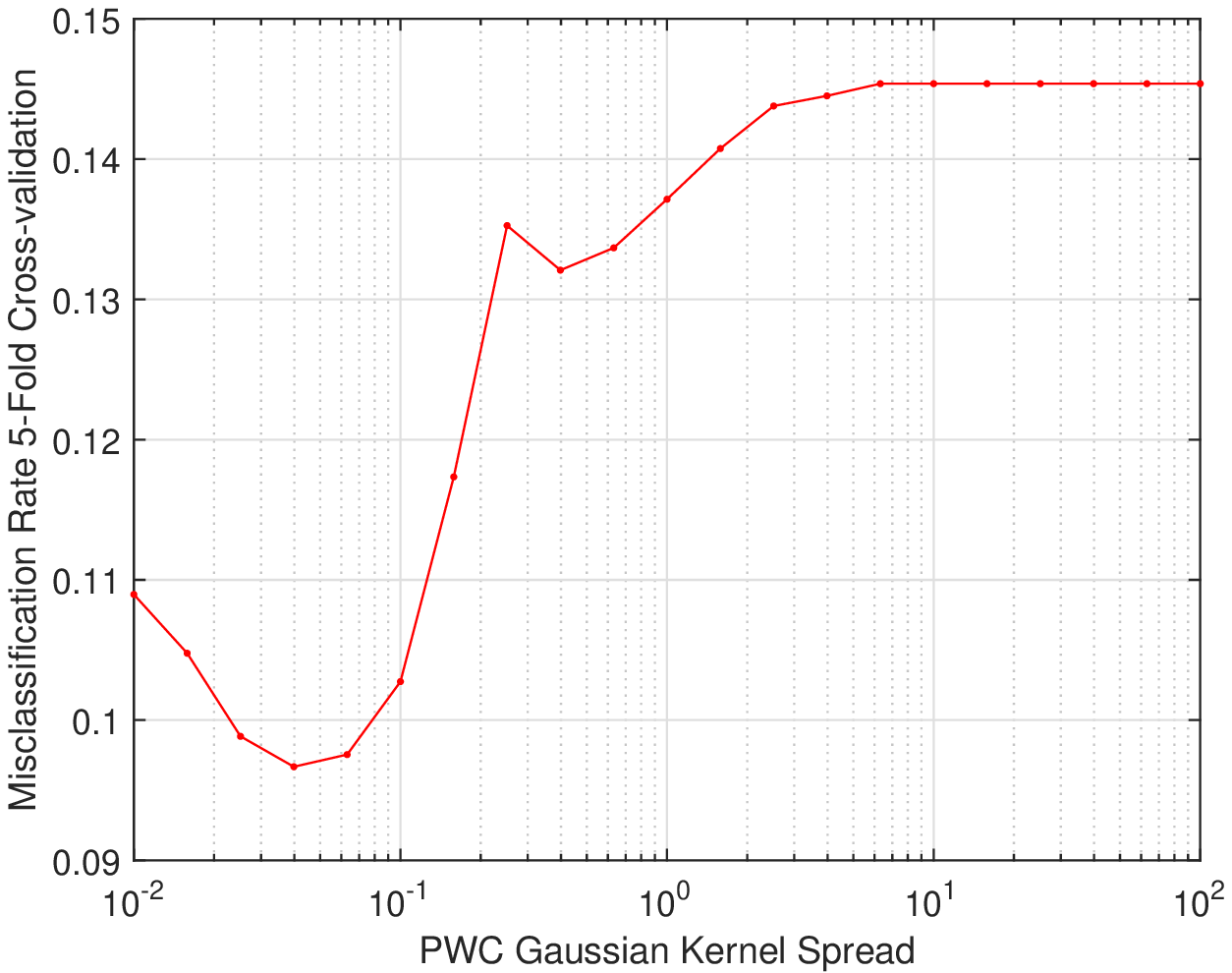}}
\par\end{centering}

\centering{}\subfloat[RBF-NN Classifier\label{fig:RBFNN_UCR}]{

\protect\includegraphics[scale=0.5]{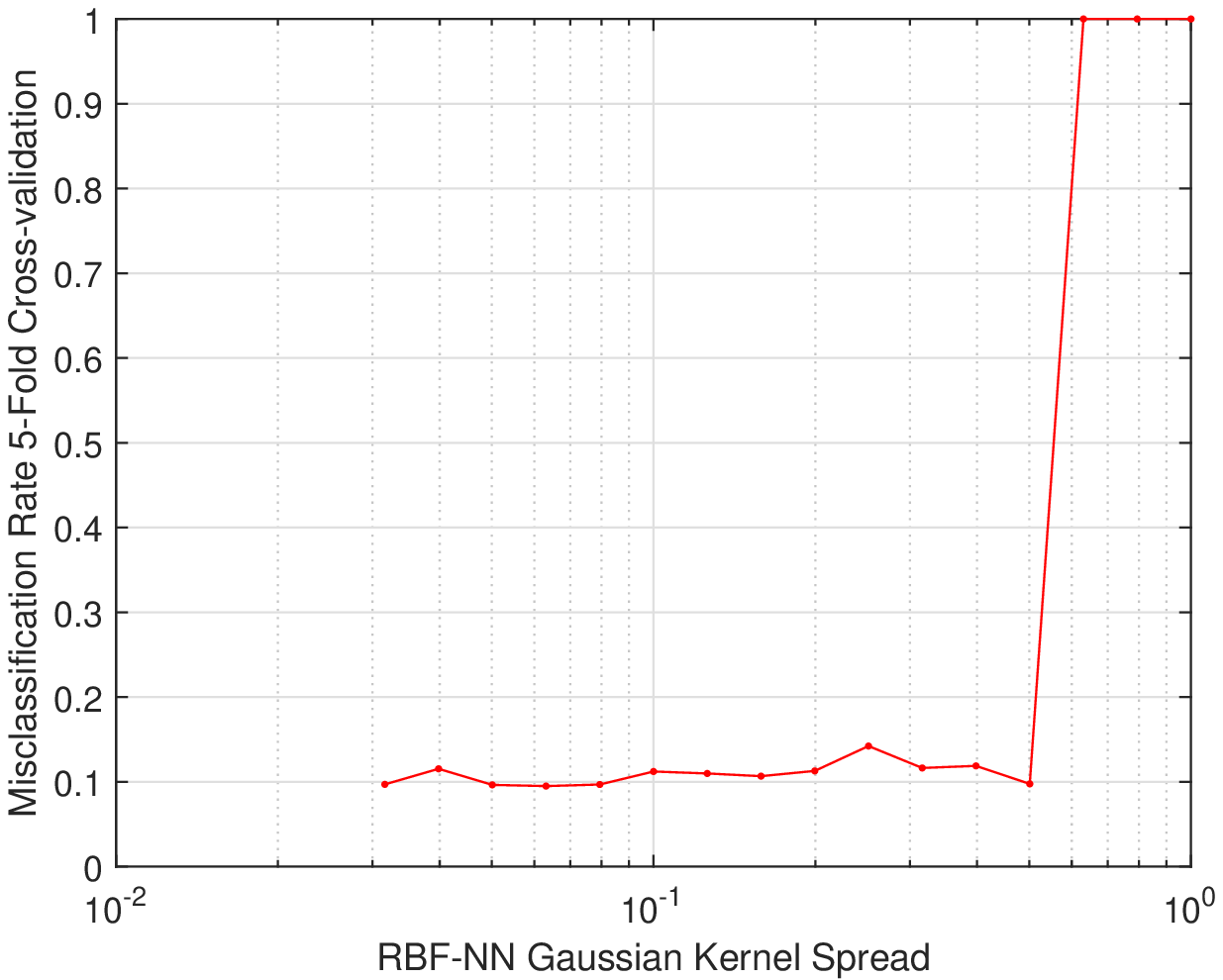}}\subfloat[Random Forest\label{fig:RandomForest_UCR}]{

\protect\includegraphics[scale=0.5]{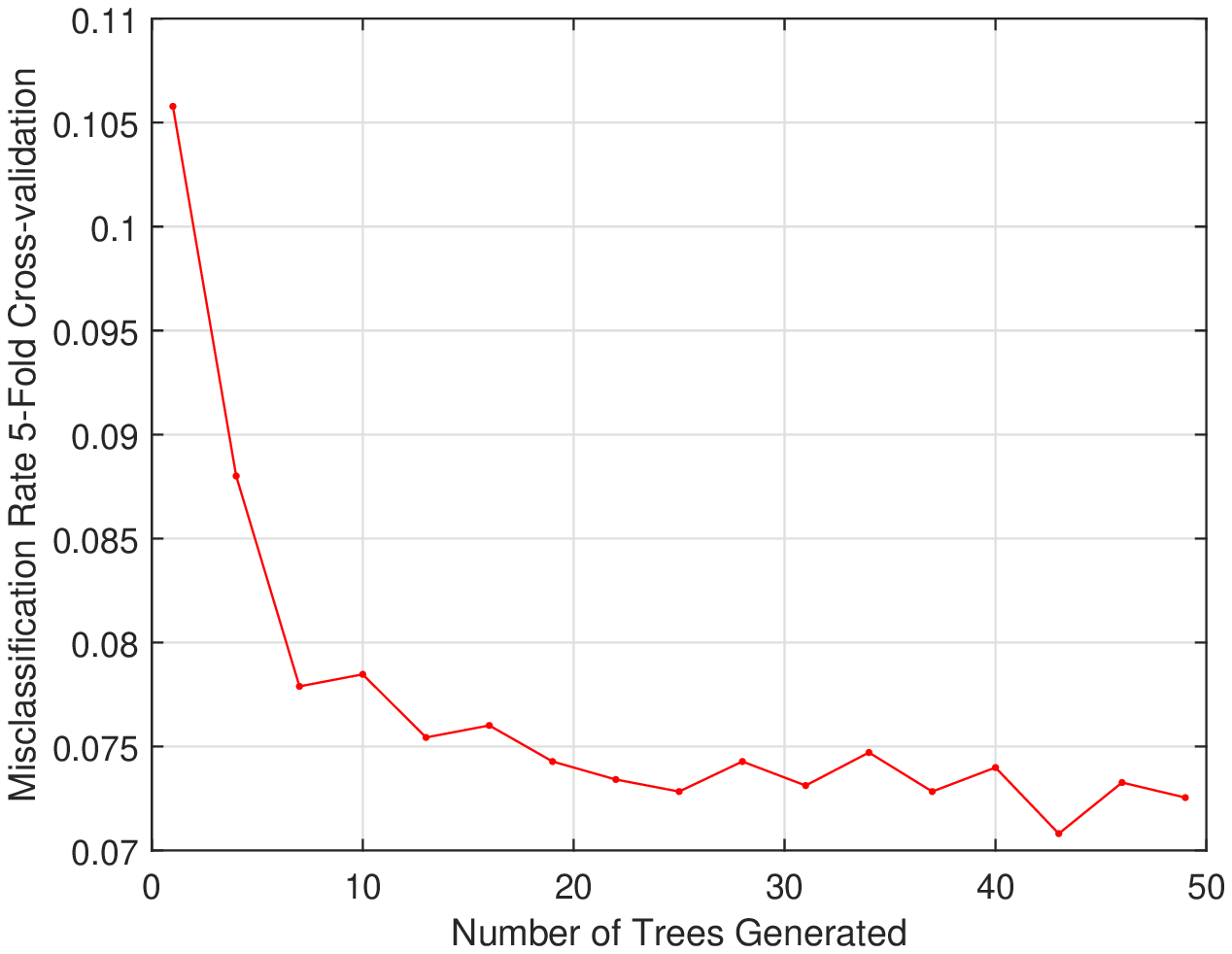}

}
\end{figure}

Each classifier is then trained only on the ECVA reduced time-domain
feature space. The resulting optimization analysis, based on the 5-fold
cross-validation is presented in Figures \ref{fig:kNN_UCR}, \ref{fig:PWC_UCR},
\ref{fig:RBFNN_UCR} and \ref{fig:RandomForest_UCR}.

Depending on the methodology used, cross-validation estimates a minimum
misclassification error of $<10\%$. The UCR website reports the following
error estimates for this dataset, note that all methods reported use
direct distance to generate a feature space (direct comparison of
curves): 1-NN Euclidean Distance ($15.1\%$), 1-NN Best Warping Window
DTW ($9.5\%$) and 1-NN DTW, no warping window ($9.3\%$). For a more
detailed comparison, the confusion matrix for each of the optimized
classifiers is presented in Tables \ref{TablekNN_UCR}, \ref{TablePWC_UCR},
\ref{TableRBFNN_UCR} and \ref{TableRF_UCR}. 

\begin{table}[H]
\protect\caption{Confusion Matrix for Classifiers Based on UCR Starlight Data }

\begin{centering}
\subfloat[1-NN\label{TablekNN_UCR}]{

\begin{tabular}{|c|c|c|c|}
\hline 
True\textbackslash{}Est & 1 & 2 & 3\tabularnewline
\hline 
\hline 
1 & 0.86 & 0.003 & 0.13\tabularnewline
\hline 
2 & 0.0 & 0.99 & 0.008\tabularnewline
\hline 
3 & 0.031 & 0.002 & 0.97\tabularnewline
\hline 
\end{tabular}}\subfloat[PWC\label{TablePWC_UCR}]{

\begin{tabular}{|c|c|c|c|}
\hline 
True\textbackslash{}Est & 1 & 2 & 3\tabularnewline
\hline 
\hline 
1 & 0.82 & 0.003 & 0.18\tabularnewline
\hline 
2 & 0.00 & 0.97 & 0.035\tabularnewline
\hline 
3 & 0.16 & 0.004 & 0.84\tabularnewline
\hline 
\end{tabular}}
\par\end{centering}

\centering{}\subfloat[RBF-NN\label{TableRBFNN_UCR}]{

\begin{tabular}{|c|c|c|c|}
\hline 
True\textbackslash{}Est & 1 & 2 & 3\tabularnewline
\hline 
\hline 
1 & 0.91 & 0.003 & 0.082\tabularnewline
\hline 
2 & 0.065 & 0.94 & 0.0\tabularnewline
\hline 
3 & 0.049 & 0.0007 & 0.95\tabularnewline
\hline 
\end{tabular}}\subfloat[RF\label{TableRF_UCR}]{

\begin{tabular}{|c|c|c|c|}
\hline 
True\textbackslash{}Est & 1 & 2 & 3\tabularnewline
\hline 
\hline 
1 & 0.91 & 0.003 & 0.082\tabularnewline
\hline 
2 & 0.0 & 0.99 & 0.005\tabularnewline
\hline 
3 & 0.004 & 0.0007 & 0.99\tabularnewline
\hline 
\end{tabular}

}
\end{table}

\subsubsection{Discussion}

The SSMM methodology presented does no worse than the 1-NN presented
by \citet{Keogh2011} and appears to provide some increase in performance.
The procedure described operates on folded data as well as unfolded
data and does not need time-warping for alignment of the waveform,
demonstrating the flexibility of the method. The procedure not only
separated out the classes outlined, but in addition found additional
clusters of similarity in the dataset. If these clusters correspond
to the sub-groupings reported by the original generating source (RRab
and RRc, etc.) is not known, as object identification is not provided
by the UCR dataset.

\subsection{Application to New Set (LINEAR)}

For the analysis of the proposed algorithm design, the LINEAR dataset
is parsed into training, cross-validation and test sets on time series
data from the LINEAR survey that has been verified, and for which
accurate photometric values are available \citep{Sesar2011,Palaversa2013}.
From the starting sample of 7194 LINEAR variables, a clean sample
of 6146 time series datasets and their associated photometric values
were used for classification. Stellar class type is limited further
to the top five most populous classes: RR lyr(ab), RR lyr (c ), Delta
Scuti / SX Phe, Contact Binaries and Algol-Like Stars with 2 Minima;
resulting in a set of 6086 observations, the distribution of stellar
classes is presented in Table \ref{TableDistribution_LINEAR}.

\begin{table}[H]
\protect\caption{Distribution of LINEAR Data Across Classes\label{TableDistribution_LINEAR}}

\begin{centering}
\begin{tabular}{|c|c|c|}
\hline 
Type & Count & Percent\tabularnewline
\hline 
\hline 
Algol & 287 & 4.7\%\tabularnewline
\hline 
Contact Binary & 1805 & 29.7\%\tabularnewline
\hline 
Delta Scuti & 68 & 1.1\%\tabularnewline
\hline 
No Variablity & 1000 & 16.4\%\tabularnewline
\hline 
RRab & 2189 & 36.0\%\tabularnewline
\hline 
RRc & 737 & 12.1\%\tabularnewline
\hline 
\end{tabular}
\par\end{centering}

\end{table}

\subsubsection{Non-Variable Artificial Data}

In support of the supervised classification algorithm, artificial
datasets have been generated and introduced into the training/testing
set. These artificial datasets are representation of stars with-out
variability. This introduction of artificial data is done for the
same reasons the training of the anomaly detection algorithm is performed:
\begin{itemize}
\item The LINEAR dataset implemented only represents five of the top (most
populous) variable star types, while at least 23 stellar variable
types are known \citep{Richards2012}, thus the class space defined
by the classes is incomplete.
\item Even if the class space was complete, studies such as \citet{Debosscher2009,Dubath2011}
have all shown that many stellar variable populations are under-sampled. 
\item Similarly, many of the studies focus on stellar variables only, and
do not include non-variable stars. While filters are often applied
to separate variable and non-variable stars (Chi-Squared specifically),
these are not necessarily perfect methods for removing non-variable
populations, and could result in an increase in false alarms. 
\end{itemize}
This artificial time series is generated with a Gaussian Random amplitude
distribution. In addition to the time-domain information randomly
generated, differential photometric information is also generated.
The differential photometric measurements used to classify the stars
are used to generate empirical distributions (histograms) of each
of the feature vectors. These histograms are turned into cumulative
distribution functions (CDFs). The artificially generated differential
photometric patterns are generated via sampling from these generated
empirical distribution functions. Sampling is performed via the Inverse
Transform method \citep{Law1991} . These artificial datasets are
treated identical in processing to the other observed waveforms.

\subsubsection{Time Domain and Differential Photometric Feature Space}

In addition to the time domain data, differential photometric data
is obtainable for the LINEAR dataset, resulting from the efforts of
large photometric surveys such as SDSS and 2MASS. These additional
features are merged with the reduced time domain feature space, resulting
in an overall feature space. For this study, the optical SDSS filters
($ugriz$) and the IR filters ($JK$) are used to generate the differential
features: $u-g$, $g-i$, $i-K$ and $J-K$. The color magnitudes
are corrected for the ISM extinction using $E(B-V)$ from the SFD
maps and the extinction curve shape from \citet{Berry2012}. In addition
to these differential color domain features, bulk time domain statistics
are also generated: $logP$ is the log of the primary period derived
from the Fourier domain space, $magMed$ is the median LINEAR magnitude,
$ampl$, $skew$ and $kurt$ are the amplitude, skewness and kurtosis
for the observed light curve distribution. These additional features
will be included for the analysis of the LINEAR dataset. See electronic
supplement (Combined LINEAR Features, Extra-Figure-CombinedLINEARFeatures.fig)
for a plot matrix of the combined feature space.

\subsubsection{Analysis}

It is assumed that the parameters that minimize the mis-classification
rate using QDA, will likewise minimize the mis-classification rate
using any of the other classification algorithms. The error resulting
from mis-classification is minimized resulting from the cross-validation,
optimizing both the kernel width associated with the slotting method
as well as the state space resolution of the symbolic alphabet. Using
the optimal parameters, a three dimensional plot (the first three
ECVA parameters) is constructed; see the electronic supplement for
the associated movie (ECVA Feature LINEAR Movie, ExtendedCanonicalVariates.mp4).
Figure \ref{fig:ECVA_LINEAR} is a plot of the first two extended
canonical variates:

\begin{figure}[H]
\protect\caption{First two Extended Canonical Variates for the Time-Domain Feature
Space \label{fig:ECVA_LINEAR}}

\begin{centering}
\includegraphics[scale=0.5]{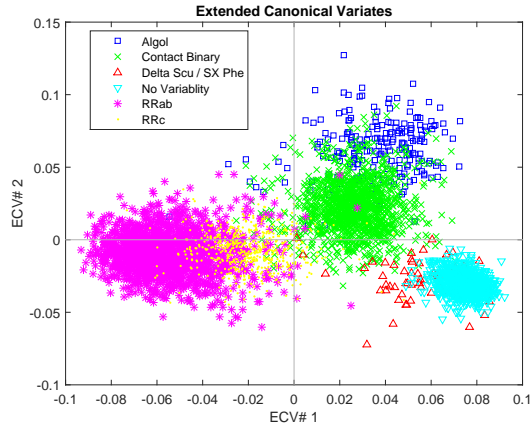}
\par\end{centering}

\end{figure}

\begin{figure}[H]
\protect\caption{Classifier Optimization for LINEAR Data}

\begin{centering}
\subfloat[Nearest Neighbor classifiers\label{fig:kNN_LINEAR}]{

\protect\includegraphics[scale=0.5]{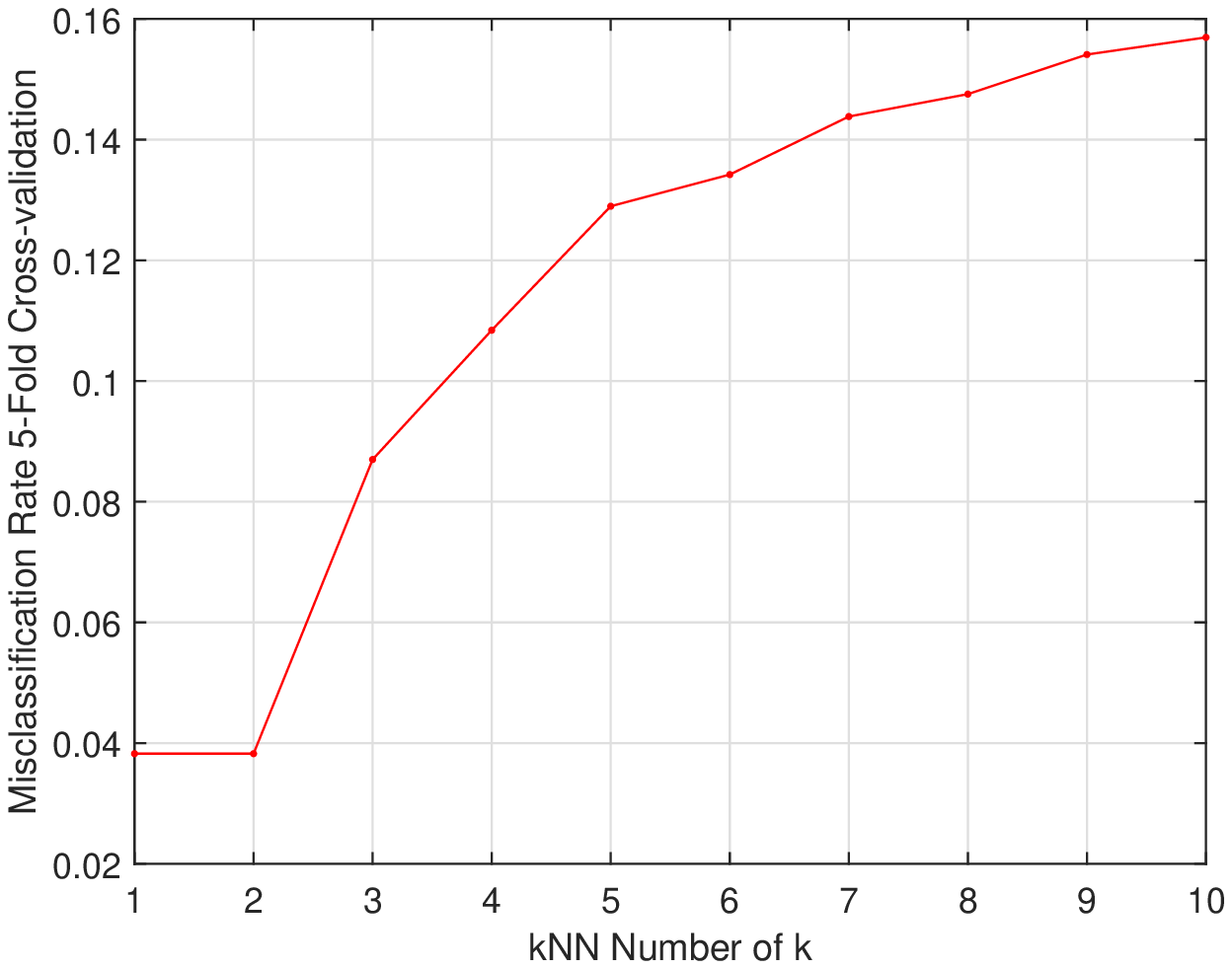}}\subfloat[Parzen window classifier\label{fig:PWC_LINEAR}]{

\protect\includegraphics[scale=0.5]{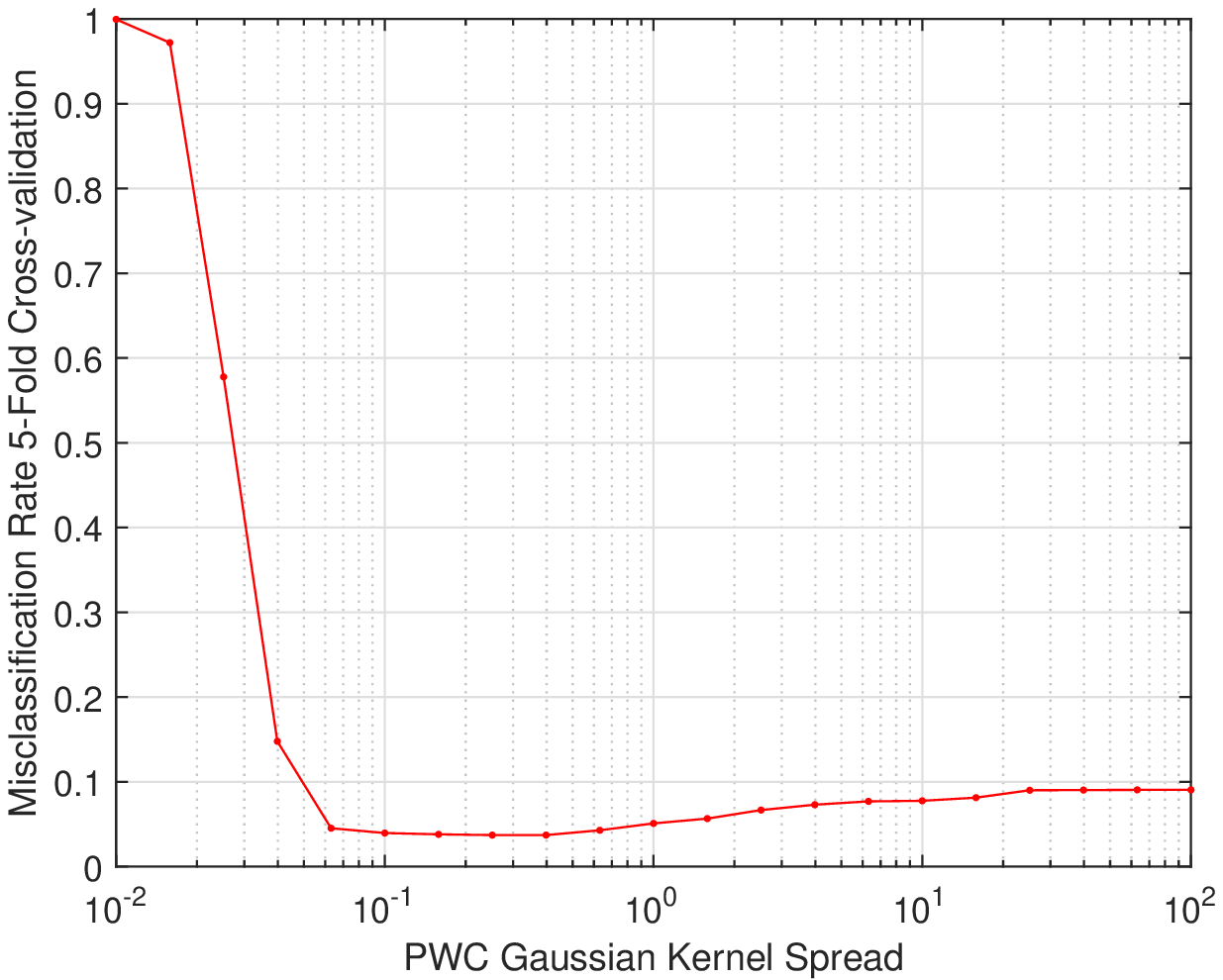}}
\par\end{centering}

\centering{}\subfloat[RBF-NN Classifier\label{fig:RBFNN_LINEAR}]{

\protect\includegraphics[scale=0.5]{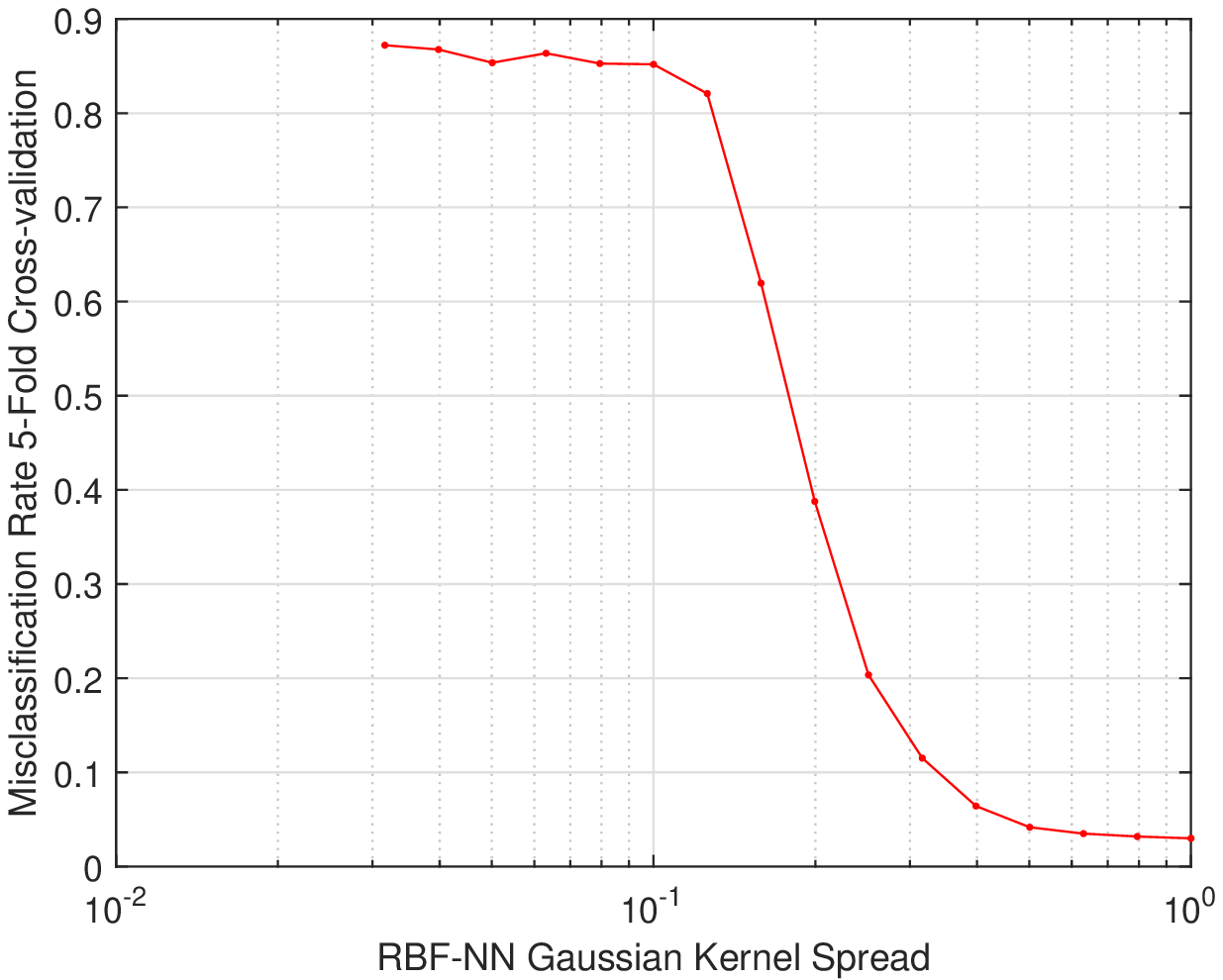}}\subfloat[Random Forest\label{fig:RandomForest_LINEAR}]{

\protect\includegraphics[scale=0.5]{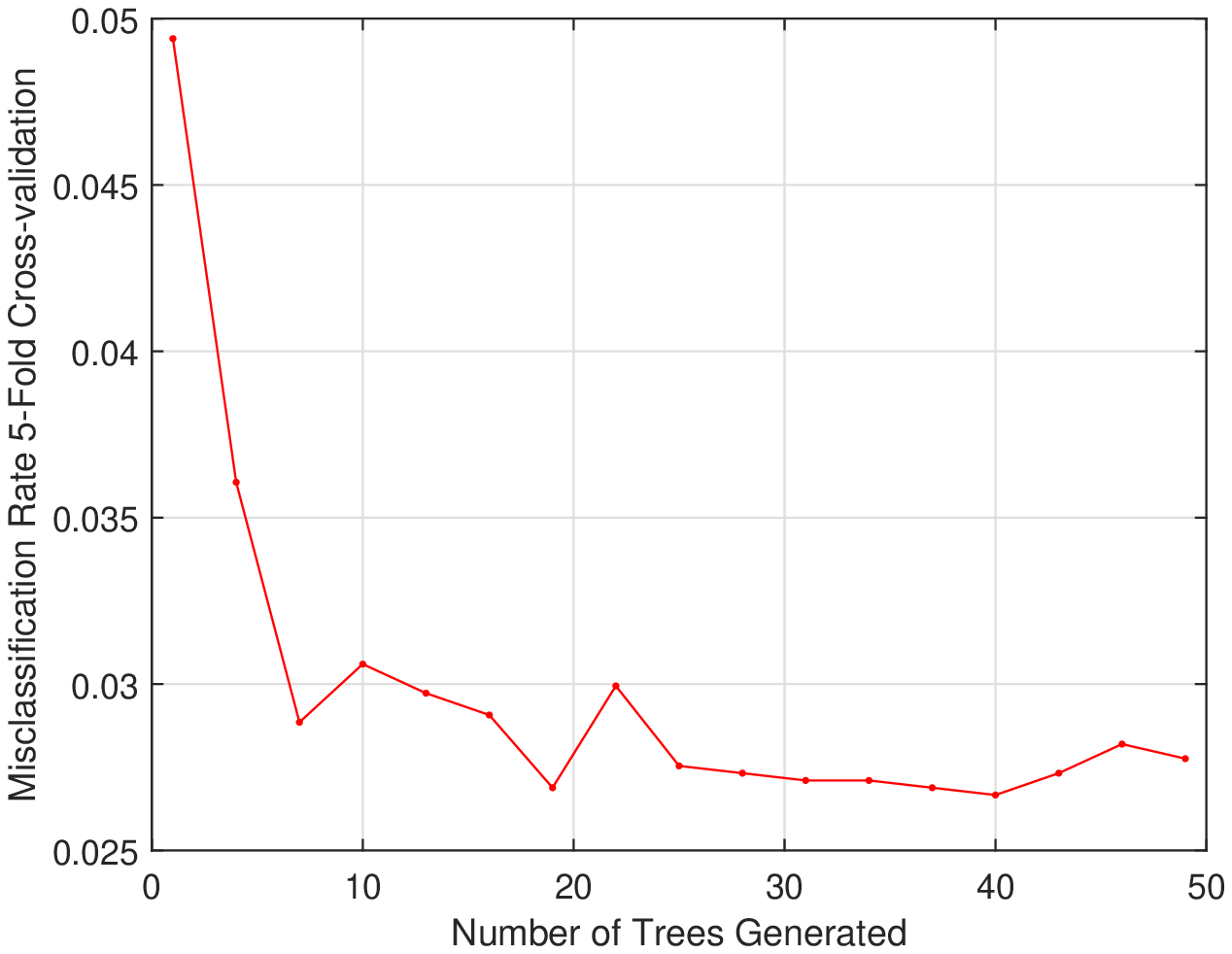}

}
\end{figure}

Based on the merged feature space, the optimal parameters for the
kNN, PWC, RBF-NN and Random Forest Classifier are generated. The error
analysis figures for each are presented in Figures \ref{fig:kNN_LINEAR},
\ref{fig:PWC_LINEAR}, \ref{fig:RBFNN_LINEAR} and \ref{fig:RandomForest_LINEAR}
respectively. 

Testing was performed on a pre-partitioned set, separate from the
training and cross-validation populations. The transformation applied
to the training and cross-validation data were also applied to the
testing data (including centering and rotating). After optimal parameters
have been found for both the resolution of the Markov Model and the
classification algorithms, the testing set is used to estimate the
confusion matrix. A confusion matrix is generated \textquotedblleft True
labels\textquotedblright{} are shown on the left column and \textquotedblleft Estimated
label\textquotedblright{} are shown on the top row (Tables \ref{TableKNN_LINEAR},
\ref{TablePWC_LINEAR}, \ref{TableRBFNN_LINEAR} and \ref{TableRF_LINEAR}).

\begin{table}[H]
\protect\caption{Confusion Matrix for Classifiers Based on UCR Starlight Data }

\begin{centering}
\subfloat[1-NN\label{TableKNN_LINEAR}]{

\begin{tabular}{|c|c|c|c|c|c|c|}
\hline 
True\textbackslash{}Est & Algol & Contact Binary & Delta Scuti & No Variation & RRab & RRc\tabularnewline
\hline 
\hline 
Algol & 0.76 & 0.20 & 0.0 & 0.0 & 0.0 & 0.04\tabularnewline
\hline 
Contact Binary & 0.03 & 0.95 & 0.005 & 0.005 & 0.01 & 0.0\tabularnewline
\hline 
Delta Scuti & 0.0 & 0.0 & 0.88 & 0.12 & 0.0 & 0.0\tabularnewline
\hline 
No Variation & 0.0 & 0.0 & 0.01 & 0.99 & 0.0 & 0.0\tabularnewline
\hline 
RRab & 0.0 & 0.005 & 0.0 & 0.0 & 0.95 & 0.045\tabularnewline
\hline 
RRc & 0.0 & 0.03 & 0.0 & 0.0 & 0.14 & 0.83\tabularnewline
\hline 
\end{tabular}}
\par\end{centering}

\begin{centering}
\subfloat[PWC\label{TablePWC_LINEAR}]{

\begin{tabular}{|c|c|c|c|c|c|c|}
\hline 
True\textbackslash{}Est & Algol & Contact Binary & Delta Scuti & No Variation & RRab & RRc\tabularnewline
\hline 
\hline 
Algol & 0.97 & 0.01 & 0.0 & 0.0 & 0.02 & 0.0\tabularnewline
\hline 
Contact Binary & 0.0 & 0.99 & 0.0 & 0.0 & 0.0 & 0.01\tabularnewline
\hline 
Delta Scuti & 0.0 & 0.0 & 0.94 & 0.06 & 0.0 & 0.0\tabularnewline
\hline 
No Variation & 0.0 & 0.0 & 0.0 & 1.0 & 0.0 & 0.0\tabularnewline
\hline 
RRab & 0.0 & 0.01 & 0.0 & 0.0 & 0.99 & 0.0\tabularnewline
\hline 
RRc & 0.0 & 0.01 & 0.0 & 0.0 & 0.0 & 0.99\tabularnewline
\hline 
\end{tabular}}
\par\end{centering}

\begin{centering}
\subfloat[RBF-NN\label{TableRBFNN_LINEAR}]{

\begin{tabular}{|c|c|c|c|c|c|c|}
\hline 
True\textbackslash{}Est & Algol & Contact Binary & Delta Scuti & No Variation & RRab & RRc\tabularnewline
\hline 
\hline 
Algol & 0.95 & 0.05 & 0.0 & 0.0 & 0.0 & 0.0\tabularnewline
\hline 
Contact Binary & 0.0 & 1.0 & 0.0 & 0.0 & 0.0 & 0.0\tabularnewline
\hline 
Delta Scuti & 0.0 & 0.0 & 0.94 & 0.06 & 0.0 & 0.0\tabularnewline
\hline 
No Variation & 0.0 & 0.0 & 0.0 & 1.0 & 0.0 & 0.0\tabularnewline
\hline 
RRab & 0.0 & 0.0 & 0.0 & 0.0 & 1.0 & 0.0\tabularnewline
\hline 
RRc & 0.0 & 0.01 & 0.0 & 0.0 & 0.0 & 0.99\tabularnewline
\hline 
\end{tabular}}
\par\end{centering}

\centering{}\subfloat[RF\label{TableRF_LINEAR}]{

\begin{tabular}{|c|c|c|c|c|c|c|}
\hline 
True\textbackslash{}Est & Algol & Contact Binary & Delta Scuti & No Variation & RRab & RRc\tabularnewline
\hline 
\hline 
Algol & 0.93 & 0.07 & 0.0 & 0.0 & 0.0 & 0.04\tabularnewline
\hline 
Contact Binary & 0.0 & 0.99 & 0.0 & 0.0 & 0.0 & 0.0\tabularnewline
\hline 
Delta Scuti & 0.0 & 0.0 & 0.94 & 0.0 & 0.0 & 0.06\tabularnewline
\hline 
No Variation & 0.0 & 0.02 & 0.0 & 0.98 & 0.0 & 0.0\tabularnewline
\hline 
RRab & 0.0 & 0.0 & 0.0 & 0.0 & 1.0 & 0.05\tabularnewline
\hline 
RRc & 0.0 & 0.0 & 0.0 & 0.0 & 0.0 & 1.0\tabularnewline
\hline 
\end{tabular}}
\end{table}

\subsubsection{Anomaly Detection}

In addition to the pattern classification algorithm outlined, the
procedure outlined includes the construction of an anomaly detector.
The pattern classification algorithm presented as part of this analysis,
partition the entire decision space based on the known class type
provided in the LINEAR dataset. The random forest, kNN, MLP and SVM
two-class classifier algorithms, there is no consideration for deviations
of patterns beyond the training set observed, i.e. absolute distance
from population centers. All of the algorithms investigated consider
relative distances, i.e. is the new pattern P closer to the class
center of B or A? Thus, despite that an anomalous pattern is observed
by a new survey, the classifier will attempt to estimate a label for
the observed star based on the labels it knows. To address this concern,
a one-class anomaly detection algorithm is implemented. 

Anomaly Detection and Novelty Detection methods are descriptions of
similar processes with the same intent, i.e., the detection of new
observations outside of the class space established by training. These
methods have been proposed for stellar variable implementations prior
to this analysis \citep{Protopapas2006}. \citet{Tax2001} and \citet{Tax2003}
outline the implementation of a number of classifiers for One-Class
(OC) classification, i.e., novel or anomaly detection. Here, the PWC
algorithm (described earlier) is transformed into an OC anomaly detection
algorithm. The result is the \textquotedblleft lassoing\textquotedblright{}
or dynamic encompassing of known data patterns. The lasso boundary
represents the division between known (previously observed) regions
of feature space and unknown (not-previously observed) regions. New
patterns observed with feature vectors occurring in this unknown region
are considered anomalies or patterns without support, and the estimated
labels returned from the supervised classification algorithms should
be questioned, despite the associated posterior probability of the
label estimate. This paper implements the DD Toolbox designed by Tax
and implements the PR toolbox \citep{Duin2007}. The resulting error
curve generated from the cross-validation of the PWC-OC algorithm
resembles a threshold model (probit), the point which minimizes the
error and minimizes the kernel width is found (Figure \ref{fig:OCPWC_LINEAR}).

\begin{flushleft}
\begin{figure}[H]
\protect\caption{OC-PWC Kernel Width Optimization for LINEAR Data \label{fig:OCPWC_LINEAR}}

\centering{}\includegraphics[scale=0.5]{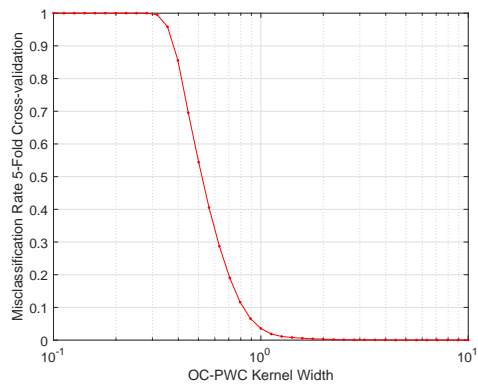}
\end{figure}

\par\end{flushleft}

This point (minimization of error and kernel width) is the optimal
kernel width (2.5). Estimated mis-classification rate of the detector
is determined via evaluation of the testing set and found to be $0.067\%$.

\subsubsection{Discussion}

Given only time series data (no differential photometric data), for
the classes and the LINEAR observations made (resolution of amplitude
and frequency rate of observations) a $\sim6\%$ mis-classification
rate with a very basic (QDA) classifier is found. Further performance
improvement is expected if other, more general, classifiers were used.
Kernel width of the slots used to account for irregular sampling and
state space resolution are major factors in performance. There exists
for our data, a point of optimal performance with respect to the kernel
width and state space resolution, that best separates the classes
observed. With the addition of differential photometric data, the
mis-classification rate is reduced by another $\sim2\%$, and results
in a nearly separable class space, depending on the methodology used
to determine the estimated class. An anomaly detection algorithm is
trained and tested on the time series data and differential photometric
data. An expected mis-classification rate of $\sim0.07\%$ is found.

\section{Conclusions}

The Slotted Symbolic Markov Modeling (SSMM) methodology developed
has been able to generate a feature space which separates variable
stars by class (supervised classification). This methodology has the
benefit of being able to accommodate irregular sampling rates, dropouts
and some degree of time-domain variance. It also provides a fairly
simple methodology for feature space generation, necessary for classification.
One of the major advantages of the methodology used is that a signature
pattern (the transition state model) is generated and updated with
new observations. The transition frequency matrix for each star is
accumulated, given new observations, and the probability transition
matrix is re-estimated. The methodology's ability to perform is based
on the input data sampling rate, photometric error and most importantly
the uniqueness of the time-domain patterns expressed by variable stars
of interest.

The analysis presented has demonstrated the SSMM methodology performance
is comparable to the UCR baseline performance analysis, if not slightly
better. In addition, the translation of the feature space has demonstrated
that the original suggestion of three classes might not be correct;
a number of additional clusters are revealed as are some potential
mis-classifications in the training set. The performance of four separate
classifiers trained on the UCR dataset is examined. It has been shown
that the methodology presented is comparable to direct distance methods
(UCR base line). It is also shown that the methodology presented is
more flexible. The LINEAR dataset provides more opportunity to demonstrate
the proposed methodology. The larger class space, unevenly sampled
data with dropouts and differential photometric data all provide additional
challenges to be addressed. After optimization, the mis-classification
rate is roughly $\sim4\%$, depending on the classifier implemented.
An anomaly detection algorithm is trained and tested on the time series
data and differential photometric data as well, with an expected mis-classification
rate of $\sim0.07\%$. The effort represents the construction of a
supervised classification algorithm.

\subsection{Future Research}

Further research is outlined in three main focus topics: dataset improvement,
methodology improvement, simulation/performance analysis. The limited
dataset and class space used for this study is known. Future efforts
will include a more complete class space, as well as more data to
support under-represented class types. Specifically datasets such
as the Catalina Real Time Transient Survey \citep{Drake2009}, will
provide greater depth and completeness as a prelude to the data sets
that will be available from the Panoramic Survey Telescope \& Rapid
Response System and the Large Synoptic Survey Telescope (LSST). 

In addition to improving the underlying training data used, the methodology
outline will also be researched to determine if more optimal methods
are available. Exploring the effects of variable size state space
for the translation could potentially yield performance improvements,
as could a comparison of slotting methods (e.g. box slots vs. Gaussian
slots vs. other kernels or weighting schemes). Likewise, implementations
beyond supervised classification (e.g., unsupervised classification)
were not explored as part of this analysis. How the feature space
outlined in this analysis would lend itself to clustering or expectation-maximization
algorithms is yet to be determined.

In a future paper, how sampling rates and photometric errors affect
the ability to represent the underlying time-domain functionality
using synthetic time-domain signals will be explored. Simulation of
the expected time domain signals will allow for an estimation of performance
of other spectral methods (DWT/DFT for irregular sampling), which
will intern allow for and understanding of the benefits and drawbacks
of each methodology, relative to both class type and observational
conditions. This type of analysis would require the modeling and development
of synthetic stellar variable functions to produce reasonable (and
varied) time domain signature.

\subsection{Acknowledgments}

The authors are grateful for valuable discussion with Stephen Wiechecki-Vergara
and Hakeem Oluseyi. Research was partially supported by Vencore, Inc.
The LINEAR program is sponsored by the National Aeronautics and Space
Administration (NRA Nos. NNH09ZDA001N, 09-NEOO09-0010) and the United
States Air Force under Air Force Contract FA8721-05-C-0002

\section{References}

\bibliographystyle{plain}
\bibliography{SPAAMRef}

\end{document}